# Electronic structure study of $TiO_2$ polymorphs, evaluation of formic acid adsorption on dry (001) and (100) $TiO_2$(B) facets by DFT calculations

Luciana Fernández-Werner, Ricardo Faccio*, Helena Pardo and Álvaro W. Mombrú

Cryssmat-Lab, DETEMA, Facultad de Química, UdelaR

Centro NanoMat, DETEMA, Polo Tecnológico de Pando, Facultad de Química, UdelaR

CINQUIFIMA, Espacio Interdisciplinario, UdelaR

Montevideo, Uruguay.

(*) rfaccio@fq.edu.uy

**Abstract**

Structural and electronic properties comparison between anatase, rutile and monoclinic $TiO_2$(B) titania polymorphs in bulk and related nanostructures simulated by two-dimensional slabs has been performed by means of density functional theory calculations. $TiO_2$(B) was included due to experimental results where it appears as a metastable phase during thermal annealing of low content sodium layered titanate nanostructures obtained via hydrothermal synthesis. Band gaps and surface energies are discussed confronting results obtained by Plane Waves (PW) and Localized Basis Set methods (LCNAO), observing an excellent agreement between both methodologies when considering floating basis to calculate surfaces energies using LCAO. The computed formation energy for the $TiO_2$(B) (001) slab, as has been previously reported by Vittadini A. *et al.*, J. Phys. Chem. C **2009**, 113 (44), 18973-18977, is in the order of the one calculated for anatasa (101) which represents one of the most stable $TiO_2$ surfaces. The adsorption of formic acid on (001) and (100) - less stable - $TiO_2$(B) dry facets was investigated as the first step for further evaluation of $TiO_2$(B) nanostructures for dye sensitized solar cells (DSSC) applications. The results show that the bidentate dissociated configuration is the most stable for both surfaces being the computed adsorption energy 0.76 eV and 1.35 eV for (001) and (100) respectively. These results differ from what was previously found for dry anatasa (101) where monodentate coordination prevails. The evaluated reactivity of $TiO_2$(B) slabs is comparable with anatase and indicates that it could be a good adsorbent for common dyes used for Dye Sensitized Solar Cells purposes.

## 1. Introduction

$TiO_2$ has been exhaustively investigated due to its unique properties and wide range of well known novel applications [1]. Most of the work has been focusing on anatase and rutile polymorphs owing to their comparative higher natural abundance, followed by brookite phase. Recently, $TiO_2$(B) polymorph has gained more attention due to its potential applications, mainly in rechargeable lithium ion batteries [2,3]but also in dye sensitized solar cells (DSSCs) [4,5] and supercapacitors [6].

$TiO_2$(B) was first synthesized by Marchand *et al.* [7], its chemical structure presents the host framework of the bronze $Na_xTiO_2$, exhibiting continuous channels and lower density than the other mentioned polymorphs. It can be easily obtained by soft chemical routes via proton exchange and subsequently dehydratation of layered titanates as was then investigated by Feist and coworkers [8,9]. It was later observed by several groups during the annealing of samples obtained performing alkali hydrothermal synthesis, first introduced by the Kasuga *et al.* [10], of titania derived nanotubes [11]. Using this approach there were synthesized different $TiO_2$(B) nanostructural morphologies as nanoparticles [5], nanowires [2,3,6], nanoribbons [4] and nanotubes [12]. Regarding their performance on Dye Sensitized Solar Cells (DSSC), at the best of our knowledge, only a few evaluations have been reported [4,5] showing comparative results.

Some theoretical studies have also been carried out about this phase. Bulk structural, electronic and vibrational properties have been investigated by Ben Yahia M. and coworkers by means of first-principles density functional theory calculations [13]. Surface reconstruction and structural stability have been studied by Vittadini *et al.* [14]. In addition, Liu W. *et al.* have studied water adsorption on $TiO_2$(B) (001) and (100) facets [15]. Moreover, lithium transport has been evaluated by Arrouvel *et al.* [16].

The presence of $TiO_2$(B) during the synthesis of titania nanostructures, showing its comparatively stability with anatase at the nanoscale, together with the relatively low investigation, motivates this study in which we intent to contribute to the knowledge about understanding of this polymorph. Furthermore, as far as we know, there are no theoretical reports on its optical properties regarding with its potential application in DSSC. As a first approach, the information about HCOOH adsorption configurations provides relevant data about the geometry of interaction with ruthenium based complexes which have carboxylic groups as bending groups, extensively used as dyes for DSSC.

The present work shows a systematic structural and electronic properties study for bulk polymorphs of $TiO_2$ and related nanostructures simulated by different two-dimensional slab models, including $TiO_2$(B) (TB) in addition to rutile (R) and anatase (A). The reported energy gaps and surfaces energies are discussed in terms of two different approaches using Plane Waves (PW) and Linear Combination of Atomic Orbitals (LCAO) methods. Then we present the analysis of the interaction of TB (001) and (100) facets with formic acid molecules comparing the adsorption energies for several different configurations including non dissociative and dissociative chemisorption. The aim of this paper is to provide more information concerning the electronic structure, stability and reactivity of this polymorph, in particular thinking about its application as semiconductor in DSSC.

## 2. Computational details and methodology

### *2.1 $TiO_2$ structural and electronic optimization*

An ab initio Density Functional Theory [17] study of structural and electronic properties for the different $TiO_2$ polymorphs bulks and surfaces was performed. Particularly, it was used DFT xc-GGA using SIESTA code [18], Quantum-ESPRESSO [19] and WIEN2k [20].

SIESTA calculations were carried out for the structural optimization of bulk and slab models. These simulations adopt a linear combination of numerical localized atomic-orbital basis sets for the description of valence electrons and norm-conserving non-local pseudopotentials for the atomic core. We selected a split-valence double-ζ basis set with polarization orbitals for Ti and O atoms. The pseudopotentials were constructed using the Trouiller and Martins scheme [21] which describes the interaction between the valence electrons and atomic core. We selected a split-valence double-ζ basis set with polarization orbitals for all the carbon atoms. The extension of the orbitals is determined by cutoff radii, as obtained from an energy shift of 50 meV due to the localization. The total energy was calculated within the Perdew–Burke–Ernzerhof (PBE) form of the generalized gradient approximation GGA xc-potential [22]. The real-space grid used to represent the charge density and wavefunctions was the equivalent of that obtained from a plane-wave cutoff of 450 Ry.

After that, the optimized bulk structures were submitted for electronic structure calculation in two different plane-wave based basis set codes: WIEN2k and Quantum ESPRESSO, in order to check the reliability of SIESTA basis sets. Firstly, the band gaps for bulk polymorphs were determined by WIEN2k, an all-electron full-potential plane-wave code, that recently implemented the latest modified Becke-Johnson exchange potential+LDA-correlation [23], which allows better estimation for band gaps. In the case of this code, the wave functions are expanded in spherical harmonics into no overlapping spheres with radii $R_{MT}$, and in plane wave into the interstitial region. The $R_{MT}$ used for Ti and O were 1.02 and 0.92 Å, respectively. Integration in reciprocal space was performed using the tetrahedron method with different set of k- points listed in Table 1. The $R_{MT}K_{max}$ parameter, which controls the size of the basis set, was $R_{MT}K_{max} = 8.0$ in all the cases.

After this procedure and with a very good correspondence between optimized bulk structures performed by the three codes we generated the following slabs: Anatase A(101), Anatase A(100), Anatase A(001), Rutile R(100), Rutile R(110), Rutile R(101), $TiO_2$(B) TB(001) in one of the possible choices and $TiO_2$(B) TB(100) (see Figure 1). The slabs were constructed by transformation of the unit cell followed by the application of a vacuum region, along the non periodical direction, of about 10 Å which is sufficient to avoid significant interactions between different images of the slabs. The cell dimensions of the original bulk unit cell, along the periodical directions of the slab, were kept fixed; while the atomic positions of all the atoms were allowed to optimize.

Given the limitations of localized basis set to overcome the problem of surface energies [24], we used a pseudopotential based plane-waves basis set code such as Quantum ESPRESSO (QE). Within this code, the electron-ion interactions were described by ultrasoft pseudopotentials [25]. Valence electrons included the O 2s and 2p and the Ti 3s, 3p, 3d, and 4s shells. The electronic states were expanded in plane waves, and the energy cutoffs for the smooth part of the wave functions and the augmented density corresponded to 40 and 400 Ry, respectively. Regarding the surface energies calculated with LCAO, in order to overcome the problem mentioned above, we introduce floating basis sets corresponding to a $TiO_2$ monolayer, following the corresponding pattern on both sides of the slab for each model. It was done in order to contemplate the basis expansion when comparing the total energies with the corresponding energies of bulk structures.

The atomic positions were fully relaxed only in the case of SIESTA, using a conjugate-gradient algorithm until all forces reach the tolerance of F= 0.04 eV/Å. In addition, the unit cells were kept fixed in the plane to the experimental data, while an about 10 Å of vacuum region was added to the perpendicular direction. For the k-point sampling of the full Brillouin zone we selected Monkhorst Pack grid [26] corresponding to 6x4x1 and 15x15x1 for the smallest and biggest supercells respectively. All these parameters allow the convergence of the total energy and forces.

### *2.2 Formic acid adsorption on clean (001) and (100) $TiO_2$(B) facets*

The simulation of the molecule-surface interactions were carried out using the SIESTA code. The adsorption energies ($E_f$) were calculated according to the expression (i). $E_T$ stands for total energy calculated for the interacting $TiO_2$ slab and HCOOH molecule at the surface, $E_{ni}$ represents the total energy calculated for non interacting slab and HCOOH molecule, and |BSSE| is the basis superposition error associated with the use of localized basis. $E_{ni}$ is computed placing an isolated *trans* formic acid molecule far from the surface slab (~5Å).

$$E_f = -\left(E_T - E_{ni} + |BSSE|\right) \text{ (i)}$$

$$|BSSE| = E_{HCOOH,dist} - E_{TiO_2,ghost} + E_{TiO_2,dist} - E_{HCOOH,ghost} \text{ (ii)}$$

The |BSSE| was estimated using equation (ii). The term $E_{HCOOH,dist}$ corresponds to the energy of a single distorted HCOOH molecule as it becomes due to the interaction with the slab surface (it could be even dissociated). Analogously, $E_{TiO2,dist}$ is the total energy calculated for the $TiO_2$ slab with the structural reconstruction caused by the interaction. Finally $E_{TiO2,ghost}$ is the total energy of the distorted HCOOH molecule expanded by the floating basis set of the $TiO_2$ surface and $E_{HCOOH,ghost}$ is the total energy of the $TiO_2$ slab expanded by the floating basis set of the molecule.

During slab calculations all of the atoms were allowed to relax keeping fixed the in plane cell parameters. The real-space grid used to represent the charge density and wavefunctions was the equivalent of that obtained from a plane-wave cutoff of 500 Ry. The Brillouin zone was sampled with a 1x15x15 k-points mesh. The convergence threshold for the residual atomic force was 0.04eV/Ang. In the case of the bulk calculations the pressure tolerance in the cell optimization was 0.05GPa. These parameters allow the total energy convergence.

The thicknesses of the slabs were chosen as the minimum unit as it appears in the corresponding bulk unit cells, and then vacuum was applied in the normal direction of the selected facets. These are c*cos(β-90°) = 6.24Å for (001) and a*cos(β-90°) =

11.64Å for (100) which implies a bilayer of Ti atoms in the first case and four layers of Ti in the second one, see Figure 1.

In order to study the surface-HCOOH interaction, the molecule was situated –in *trans* conformation- on top of the layer being the initial distances of Ti(5c)-O around 2-2.2Å for each configuration, and O(2c)-H 1.7-2.0Å or 1.1-1.4 Å for non dissociated and dissociated respectively. Ti(5c) and O(2c) refers to the surface five-fold coordinated Ti and two-fold coordinated O respectively. In all the cases all the atoms were allowed to relax.

## 3. Result and discussion

### 3.1. Bulk structures

In the case of bulk structures, there was a good accordance between codes when optimizing ionic positions and cell parameters, as presented in Table 1. The energy gaps, listed in Table 2, showed the excellent agreement between plane waves and localized basis sets and within the xc-GGA, being the difference less than 5% in the three cases. One further step consisted in the introduction of the modified Becke-Johnson potential, which allowed a better accuracy in the energy gaps, which better reflects the experimental data. All of this is a confirmation of the quality of the basis set for Ti and O atoms, in the case of the localized basis set, that allow us to obtain good structural reconstructions.

As a remark, it can be seen in Figure 1 that, unlike what occurs in anatase and rutile structures, in the case of $TiO_2$(B) some of the titanium atoms have distorted square-pyramidal environment instead of octahedral. Being the computed distance Ti-O of 2.34

Å, greater than typical Ti-O single bond length (2.20-2.25Å). This is in agreement with previous observations [13,14] and, as it will be discussed later, it influences the relative stability of the different facets.

### 3.2. Slabs

The full optimization of the slabs was performed using SIESTA and QE. Due to the differences in the symmetry of the $TiO_2$ polymorphs, different numbers of atoms were considered in each supercell. This information is included in Table 3, Figure 1 and Figure 2 where the chosen cutting planes along different lattices planes are shown. As it was mentioned before, floating basis sets were added in the LCAO calculations.

Surface energies (shown in Table 3) were determined using the expression (iii), where $E_{slab}$ stands for the total energy of the slab, $N_{Ti}$ corresponds to the number of $TiO_2$ units inside the slab, $E_{TiO2\text{-}bulk}$ is the total energy of the corresponding bulk structure per $TiO_2$ unit and finally S correspond to the surface area of the supercell used for the slab.

$$E_S = \frac{E_{slab} - N_{Ti} E_{TiO_2-bulk}}{2S} \qquad \text{(iii)}$$

Confronting the surface energies is noticeably the general agreement obtained when considering the floating basis sets in LCAO for all of the studied slabs.

<u>*Anatase slabs*</u>

The surface energies observed are in agreement with the ones reported by Vittadini *et al.* [29], 0.36 $J/m^2$ for bilayer A(101), 0.63 for three-layer A(100), and 0.91 $J/m^2$ for four-layer A(001) (referring to the Ti planes as "layers"). These values are highly affected by the degree of the unsaturation of the Ti surface, accompanied by the possibility of the atomic reconstruction at the surface or even in the inner layers. In order to somehow quantify the unsaturation of a given face, the number of "broken bonds" (bb), relative to the bulk coordination (six fold coordination for titanium atoms, and three fold coordination for O atoms) can be counted per surface unit as was discussed in a previous work [30]. As it can be seen in Table 4, the surface energy follows the general tendency of unsaturation for anatasa.

Additionally, it is interesting to observe the reconstruction due to atoms displacements occurred in A(100) where the Ti atoms belonging to the inner layer move in a way that half of the Ti atoms remain in an octahedral coordination, while the others stand in a distorted hexahedral environment. This is associated with an inward relaxation of the surface oxygen atoms involved. The distance between these oxygen atoms, situated in opposite surface of the slab, changes from 3.79 Å in bulk structure (corresponding with a cell parameter of anatasa unit cell) to 3.22 Å in this model. The related O-Ti-O angle varies from 156.13° to 124.40°, see Figure 2. This extra unsaturation can reinforce the counted surface unsaturation and thus explaining the relative higher surface energy to unstauration ratio.

In A(001) surface cations rearrange in such a way that the Ti can be considered four-fold coordinated rather than five-fold coordinated, since there is an enlargement in one of the Ti-O bond. As it is shown in Figure 2, this distance is 2.31 Å in the upper surface, while in the lower surface was found 2.27 Å, being the elongated distance in opposite direction for adjacent cations. It is worth to note than even if the Ti were considered as distorted $Ti_{5c}$ instead of $Ti_{4c}$ the calculated unsaturation would still remain bigger than the others (0.140 bb/Å$^2$).

<u>Rutile slabs</u>

Regarding rutile surface energies (or slab formation energies), taking into account the number of layers used, the computed values are in agreement with those obtained by Perron *et al.* [30], which also follows the unsaturation levels (Table 5). No inner changes in coordination were found in this polymorph while the unsatured surface titanium atoms acquire distorted square pyramidal environment.

The fact that calculations gave similar results of surface energy for R(110) and R(100) is related to the low dimensionality in the vacuum direction. Since we evaluated the surface energy using ultrathin slabs, the values obtained do no reflect the ones corresponding to macroscopic films previously reported: 0.50 $J/m^2$ for R(110) and 0.69 $J/m^2$ for R(100) [30].

$TiO_2$(B) slabs

The structure and stability of $TiO_2$(B) surfaces were previously investigated by Vittadini and coworkers [14]. They found TB(001) to be the most stable slab (0.40 $J/m^2$) while TB(100) the least one (0.76 $J/m^2$). Our results, 0.35 $J/m^2$ and 0.67 $J/m^2$ respectively, are in concordance with their observation, which confirms the exceptionally low surface energy of TB(001) facet. This evidence can support the idea of obtaining $TiO_2$(B) related nanostructures experimentally by methods such as hydrothermal treatment of anatase and rutile powders.

For the unsaturation, the broken bonds were counted considering that some of Ti atoms were originally five-fold coordinated in bulk structure and some of the O atoms were two-fold coordinated in addition to the three-fold coordination. Unsaturation values obtained can not explain the relatively stability in this case (Table 6), as it was previously noticed by Liu *et al.* [15].

After optimization the bilayer TB(001) goes into a reconstruction where atoms reach a very symmetrical configuration. For example, Ti atoms which have different environment (octahedral a square pyramidal) in bulk present the same square pyramidal geometrical environment, see Figure 3. This fact acts stabilizing this structure in comparison with TB(100), even though the latter involves lower unsaturation. Note that this structure, as was pointed by Vittadini *et al.*, reminds of the one found by the same group al when relaxing anatase A(101) film with four Ti layers, which was called *"pentacoordinated nanosheet"*, and it was predicted to be the most stable $TiO_2$ nanosheet after lepidocrocite [14, 29].

On the other hand, the unsaturated Ti atoms on top of the surface allow the bonding with electron donor species, playing an important role regarding surface reactivity. Thus, many kinds of dyes could be anchored to this semiconductor surface.

Considering the computed electronic structure, the band gaps Eg=2.61 eV and Eg=2.84 eV for TB(001) and TB(100) respectively are quite proximate to Eg=2.40 eV for calculated anatase bulk. This result is in principle positive, but further work should be done in order to evaluate its performance as a semiconductor for DSSC, such as: molecular levels alignment and good anchoring between semiconductor and dye.

### 3.3. Formic Acid Adsorption on $TiO_2$(B) (001) and (100)

The formic acid adsorption, as it was mentioned before, was studied considering two crystal facets: the major exposed TB(001) surface together with the minority TB(100) surface. Due to the bigger reactivity associated with the lower stability, TB(100) was included in the present analysis since its contribution may not be negligible when evaluating the overall reactivity of $TiO_2$(B) nanoparticles.

The adsorption geometries analyzed for TB(001) facet are shown in Figure 4. We follow the notation presented by Vittadini *et al.* [31], where M denotes monodentate coordination and B refers to bidentade coordination, depending on the number of formic acid oxygens that coordinates with unsaturated surface Ti atoms. MHa is monodentate through the carbonyl group while MHb is monodentate coordination through hydroxyl group. The configurations are classified in non-dissociative and dissociative ones, the latter are indicated with underscore followed by H (M_H and BB_H).

Besides these geometries, another possibility for monodentate molecular adsorption was observed when placing HCOOH molecule initially in a BBH position. In this geometry the H atom from the carbonyl group bonds to a three coordinated O(3c) atom from the surface, see Figure 5. It is worth noting that, as *cis* conformation was not considered, the H in *trans* molecule was initially pointing towards a surface Ti atom. A rotation of the molecule was observed during the optimization until H finally faces an O atom, no change of molecule conformation occurred.

Regarding M_H dissociative monodentate coordination, it was observed that when the hydrogen of the $HCOO^-$ anion is located opposite to the slab, the coordination evolves to a BB_H type, while initially placing the H towards the slab it remains monodentate but poorly stabilized as its nearly null absorption energy reflects (Table 7).

$TiO_2$(B) TB(100) differs from previous orientation in the number of non equivalent Ti and O surface atoms. This added a new number of interaction options, see Figure 6.

Results concerning adsorption energy, bond lengths, and molecular angles are listed on Tables 7 and 8. For both facets it is observed the same result which indicates that dissociated bidentate configuration is the most stable one, followed by monodentate configuration through carbonyl group. The C-O bond lengths and C-O-C angle obtained for the isolated molecule are listed in Table 9.

The results show that formic is adsorbed in a dissociated bridging bidentate geometry on dry $TiO_2$(B) TB(001) and TB(100) surfaces. BB_H results to be the most stable configuration for both models, despite the fact that it implies a surface distortion involving the shortness of two related adjacent $Ti_{5c}$ distance (from 3.19 to 3.06 Å) in the case of TB(001), see Figure 7. The resulting molecular angle remains slightly major than in the isolated molecule. There is also an upward displacement of $O_{2c}$ bonded to the H atom together with a dislocation of the neighboring right $Ti_{5c}$, modifying the $Ti_{5c}$-$O_{2c}$-$Ti_{5c}$ angle from 146.0° to 140.8°. For TB(100) the distortion is subtler, being the contraction of about 0.06Å. In this case, the size of the supercell selected allows a unique molecule to bend as BB_H along b direction (Figure 6), which implies the saturation of the correspondingly $O_{2c}$ row. In order to evaluate this saturation, the same conformation was studied considering a supercell duplicating b parameter, and the difference observed in energy adsorption was not significant (about 0.07eV).

It is worth noting the difference with adsorption onto anatase A(101) surface on which the molecular monodentate is preferred. It can be explained observing the difference in distance between two $Ti_{5c}$ ions, which in $TiO_2$(B) TB(001) is d= 3.19Å similarly to rutile R(110) d= 2.96 Å where the BB_H configuration also prevails, these distances fit the "byte" of the formate ligand [31]. While in anatase this distance is much longer: 3.71 Å. However, we obtained the same result for $TiO_2$(B) (100) where the distance between the unsaturated Ti atoms is 3.74 Å.

Regarding the rest of the configurations, on surface TB(001) the trend in relative stability is the following: MHa_2c>MHa_3c>MHb>M_H. The difference between the two MHa options is 0.05 eV, while the monodentate through the hydroxyl group differs more significantly in 0.45 – 0.40 eV. This is in agreement with what was observed for anatase A(101) [31] and reflects the fact that the oxygen prefers two-fold coordination.

Finally, as it was mentioned before, dissociative monodentate configuration is poorly stabilized being its computed energy of 0.02 eV.

On surface TB(100), considering the most stable option of each kind of configuration, the tendency is modified to MHa>M_H>MHb. Although the spontaneous dissociation of HCOOH was not observed during the optimization, the fact that M_H is stabilized with respect to MHb could be indicating the strongest basicity of the surface oxygens in this surface. Even more, it is clear the difference in basicity between the two nonequivalent oxygen atoms, since the configuration in which H bonds to $O^*_{2c}$ is about 0.29 eV higher in energy in comparison to H when it bonds to $O_{2c}$. The same difference in energy was found between the two options in bidentade bridging geometries. On the other hand this fact can not explain the energy difference of about 0.62 eV, computed for the MHa_1 and MHa_2, since in this case the first one is the one which involves de H bonding with $O_{2c}$ and it is the most stable. Clearly what occurs is that the geometry of the surface prevents the oxygen atoms from bonding the Ti atom, where the distance between O(2)-$Ti_{5c}$ is about 3.61Å. This fact shows that this is not a monodentate coordination but it is a molecule to surface interaction mediated, possibly, by an hydrogen bond.

Finally, the difference in magnitude observed for the same configuration changing from TB(001) to TB(100) surface, indicates that the increase of energy in the case of the latter allows us to demonstrates the reactivity of this facet, which is in accordance with its lower stability, as discussed before. It is worth noting that our simulation neglects the effect of the hydration of the surface, which can affect the relatively stability among configurations due to the formation of coordination geometries, as was observed by Vittadini *et al.* in the case of anatasa A(101) [31].

As a final remark, and in order to validate our methodology, the adsorption energies of formic acid in anatase A(101) for MHa and BB_H configurations (Figure 8) were computed. In this case, in addition to the LCAO, PW based calculations were also performed. Although we found a relatively good agreement regarding the relative stability between the studied configurations, a noticeable difference in the reported value of the adsorption energy was found. In a previous work the reported formation energy corresponds to [31] 0.92 eV for monodentate molecular adsorption (MHa_2) and 0.68 eV for bidentate bridging dissociative configuration, see Table 10.

Due to the observed differences among energies obtained when using LCAO comparing with PW, additional calculations were performed considering fixed floating basis corresponding to a monolayer of $TiO_2$ onto the slab surfaces (LCAO_FB). This allowed better relaxation of the surfaces increasing the absorption energy values. Although this method seems to yield more accurate estimation, the general tendency is maintained considering a tolerance of 0.05eV. Thus, we strongly support that the fact of not using floating basis for the previous discussions of HCOOH adsorption onto TB do not affect the general conclusions about relative stability.

## 4. Conclusions

Using both methods we can conclude about the quality of the basis used for Ti and O in the case of SIESTA, obtaining very good agreement between both methodologies for structural parameters and gap energies. The use of modified Becke-Jhonson functional to evaluate the energy gaps in bulk structures yielded to a better concordance with experimental values.

The use of floating basis set for LCAO methods allowed to obtain results in excellent agreement with those obtained when using PW. These values stress the low energy formation of $TiO_2$(B) TB(001) bilayers, comparable with anatase A(101) which points out the relative stability of this phase at the nanoscale. Thus, it results interesting to study this structure as a possible building block for other high aspect nanostructures such as nanotubes and nanowires.

The comparison between the two facets TB(001) and TB(100) showed the less stability of the second one accompanied by a greater reactivity, as expected. However, in both structures dissociative bidentate bridging configuration seems to be the most stable, onto both clean surfaces. These facts could motivate further work regarding the influence of solvation on the surface reconstruction.

Regarding the reactivity of the semiconductor surfaces, it seems to be adequate for bidentate anchoring of N3 and derived dyes. Considering both the dissociative bidentate bridging configuration of HCOOH and also the optimal distance between the pairs of $Ti_{5c}$ (9.7Å in TB(001)) -which fits the distance in N3 complex (~9.8Å)-.

Finally, we proved the potentiality of mixing different schemes in order to overcome relatively big problems expanding the possibilities of the localized basis sets, without reaching the huge computational demand of plane waves approaches.

**Acknowledgments**

The authors gratefully acknowledge PEDECIBA, CSIC, ANII -Uruguayan organizations- and Uruguay-INNOVA UE-ANII project for financial support.

## FIGURE CAPTIONS

**Figure 1**.- TiO2(B) unit cell showing the two cutting options along the direction (001) (a), TB(001) chosen slab with c as vacuum direction (b), and TB(100) slab with a as vacuum direction (c).

**Figure 2.**– Investigated anatase and rutile slabs after optimizing, being the vacuum direction vertical in all the cases. The figures are presented as supercells for better visualization. The coordination of surface atoms are indicated, nc refers to n-fold coordination.

**Figure 3.-** Representation of bulk $TiO_2$(B) (a) , TB(100) (b), and TB(001) (c) y (d). Ti and Ti* represent octahedral and square pyramidal coordination respectively in bulk structure. In the case of slabs the same notation refers both to $Ti_{5c}$, in which Ti belonged to octahedral environment and Ti* was already five-fold coordinated in bulk.

**Figure 4.**– Adsorption geometries which remain stable during the optimization for the TB(001) slab.

**Figure 5.**– Monodentate adsorption through carbonyl group involving a surface $O_{3c}$.

**Figure 6.**– Adsorption geometries which remain stable during the optimization for the TB(001) slab.

**Figure 7.**– Surface distortion induced in BB_H configuration.

**Figure 8.**– Adsorption geometries evaluated for anatase A(101) slab.

## TABLE CAPTIONS

**Table 1.-** Cell parameters obtained for the two basis set: PW and LCAO

**Table 2.-** Energy Gaps (eV) for bulk structures according to PW and LCAO Basis Set

**Table 3.-** Surface energies for the simulated slabs.

**Table 4.-** Surface unsaturation for anatase slabs.

**Table 5**.- Surface unsaturation for rutile slabs.

**Table 6**.- Surface unsaturation for $TiO_2$(B) slabs.

**Table 7.-** Adsorption energies, bond distances and bond angles for the different adsorption configurations on **TB(001)** surface.

**Table 8**.- Adsorption energies, bond distances and bond angles for the different adsorption configurations on **TB(100)** surface.

**Table 9.-** HCOOH bond distances and angle for isolated molecule (far from the surface).

**Table 10.-** Adsorption energies (eV), bond distances and bond angles for the different adsorption configurations on **anatase A(101)** surface.

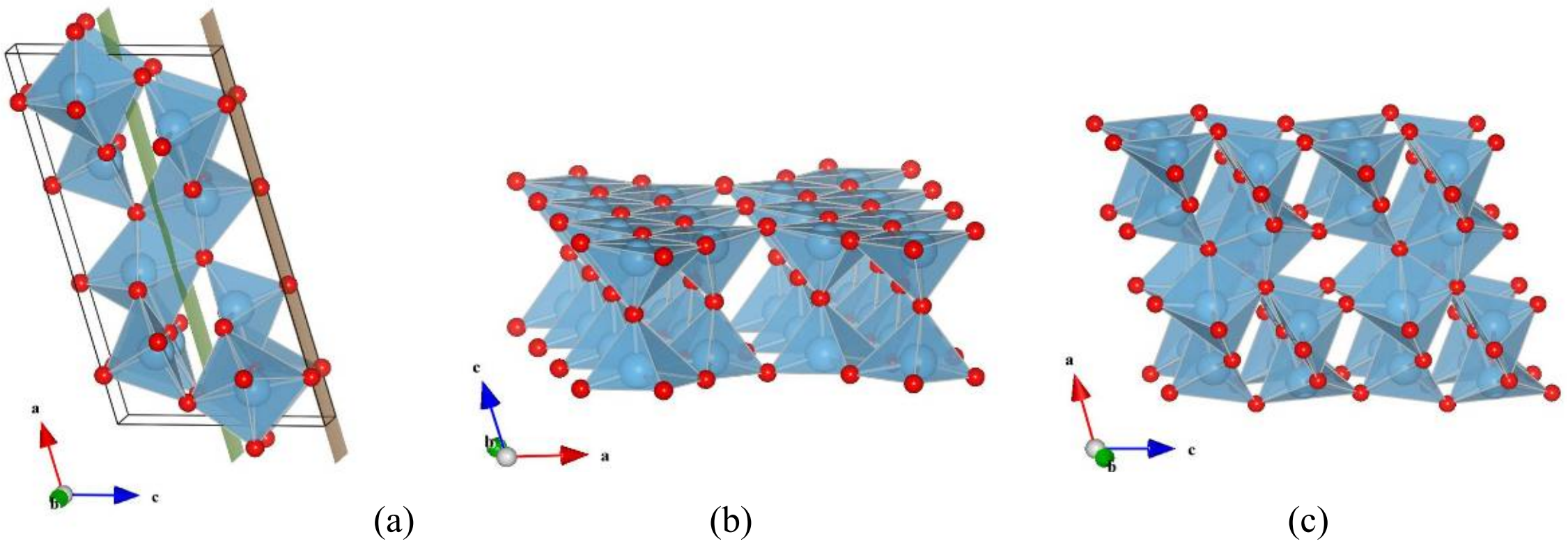

**Figure 1**

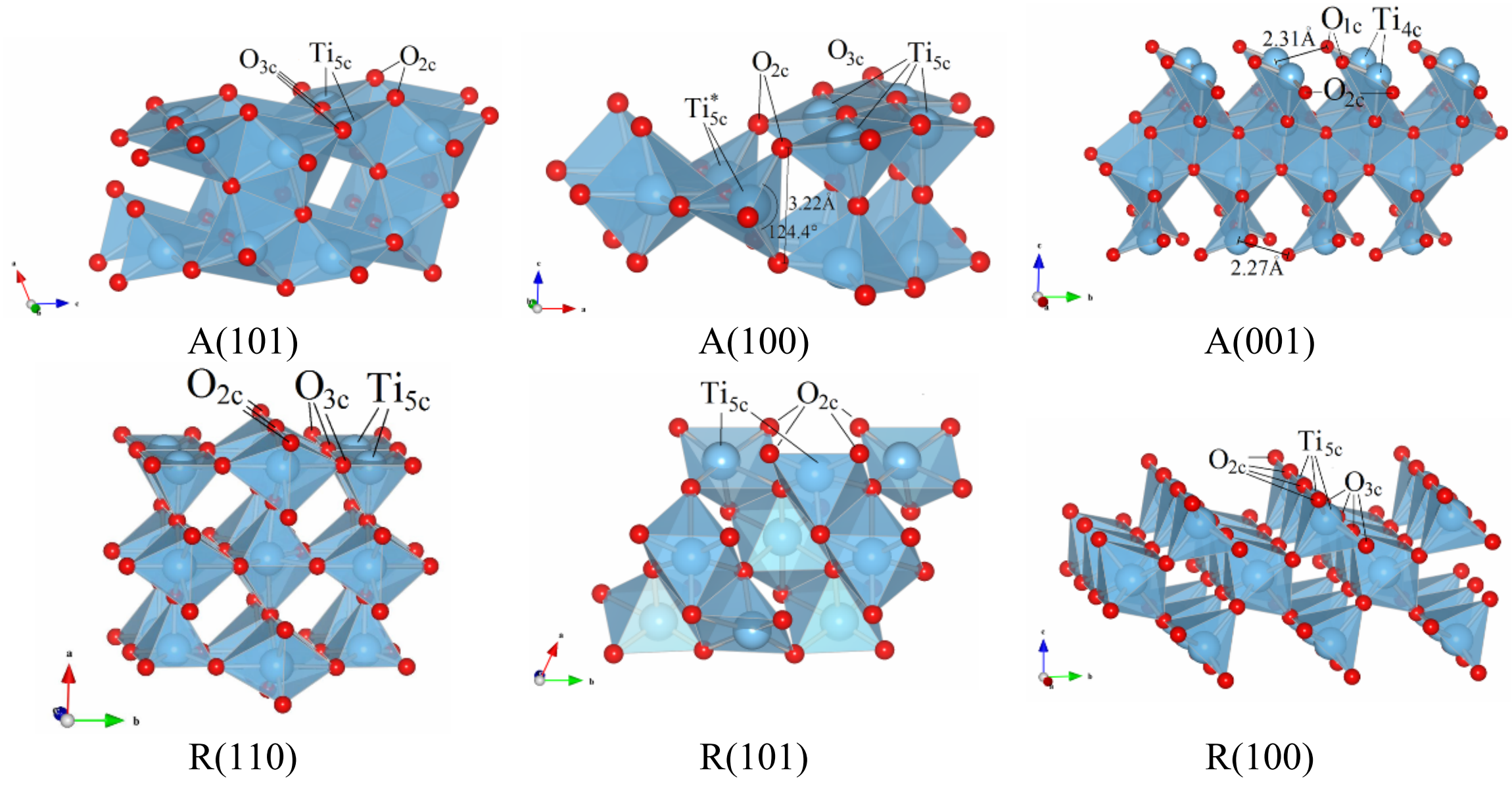

**Figure 2**

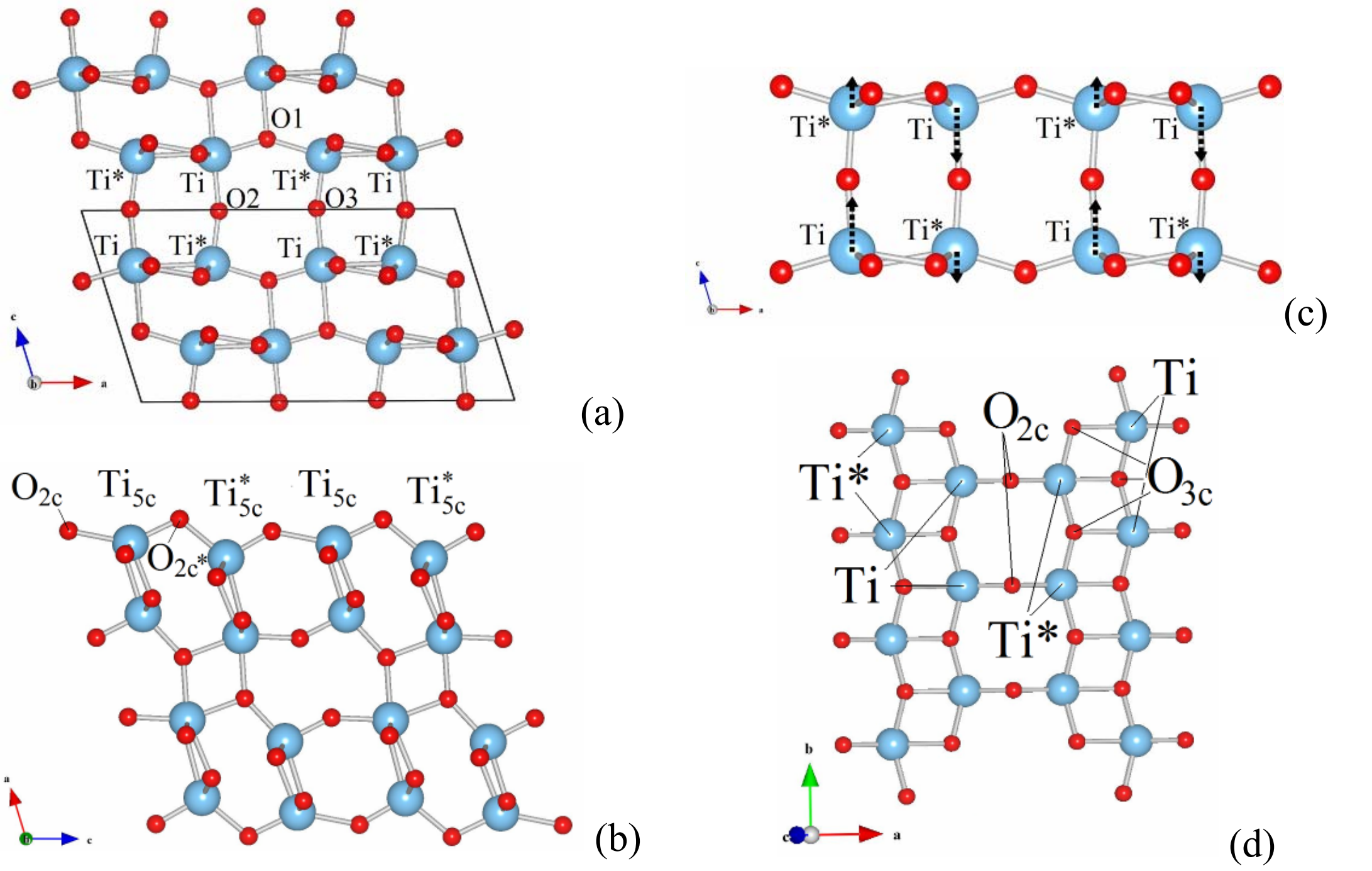

**Figure 3**

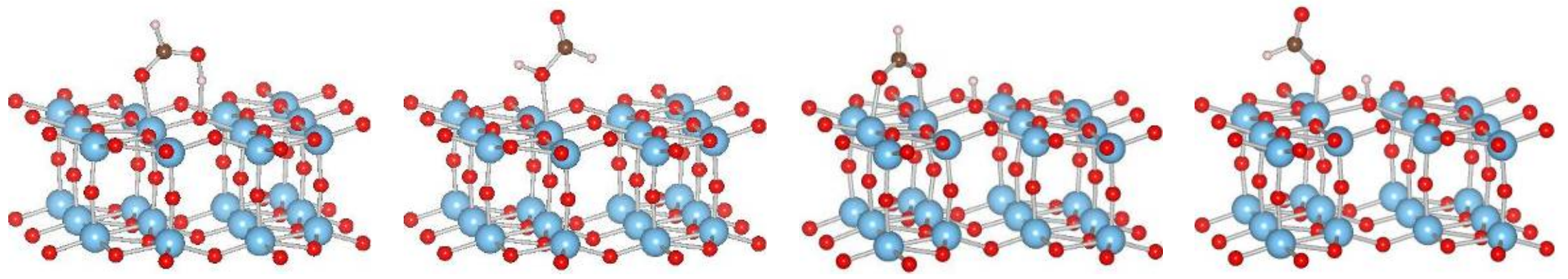

MHa MHb BB_H M_H

**Figure 4**

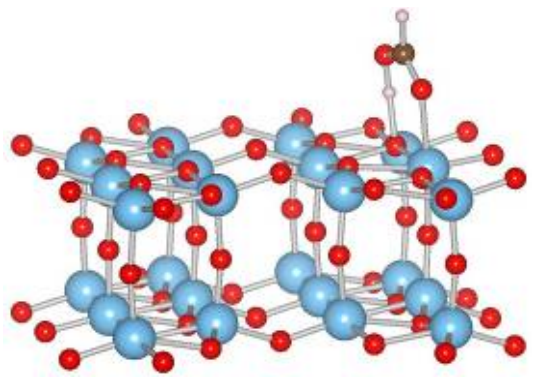

MHa_3c

**Figure 5**

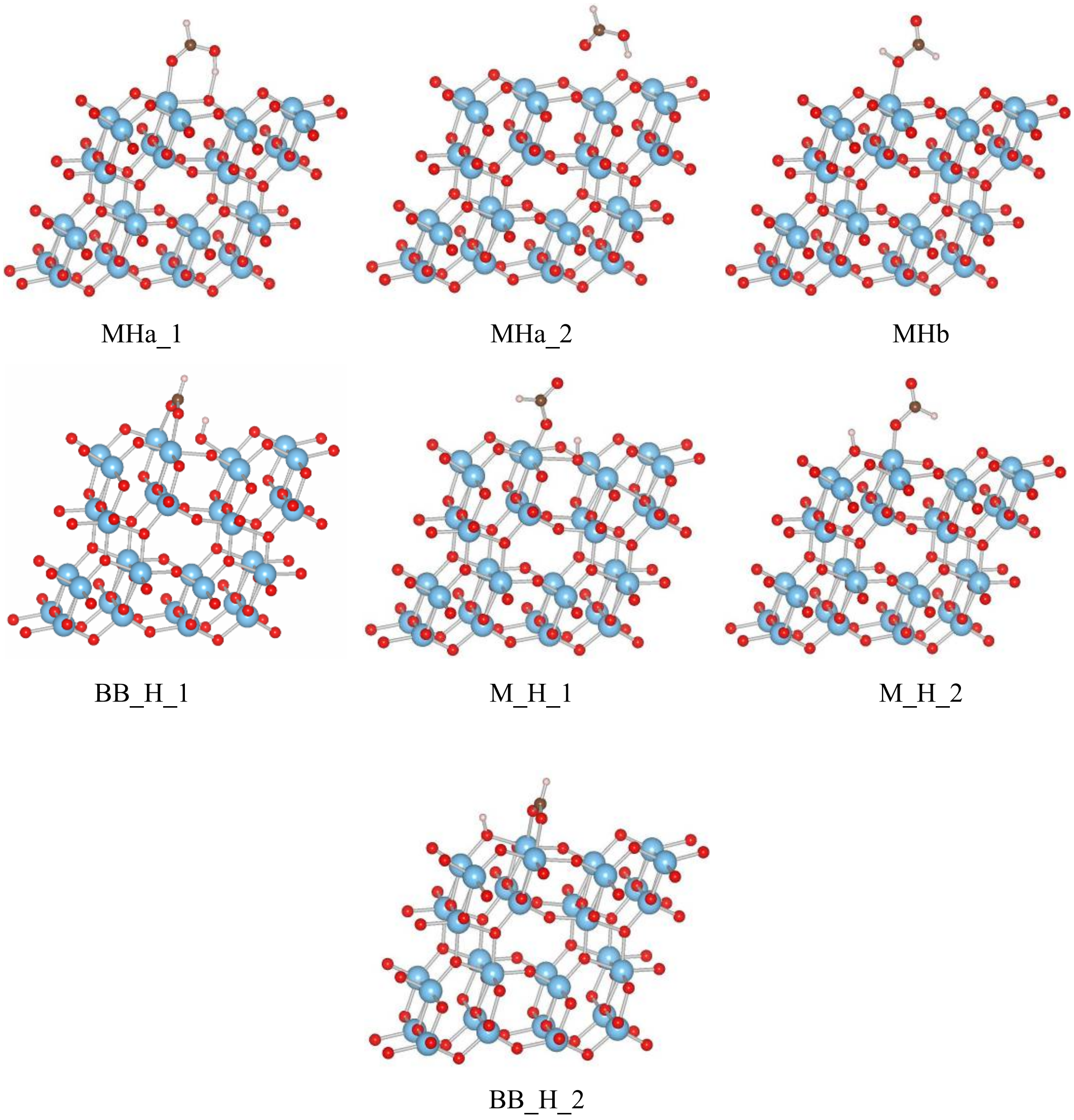

**Figure 6**

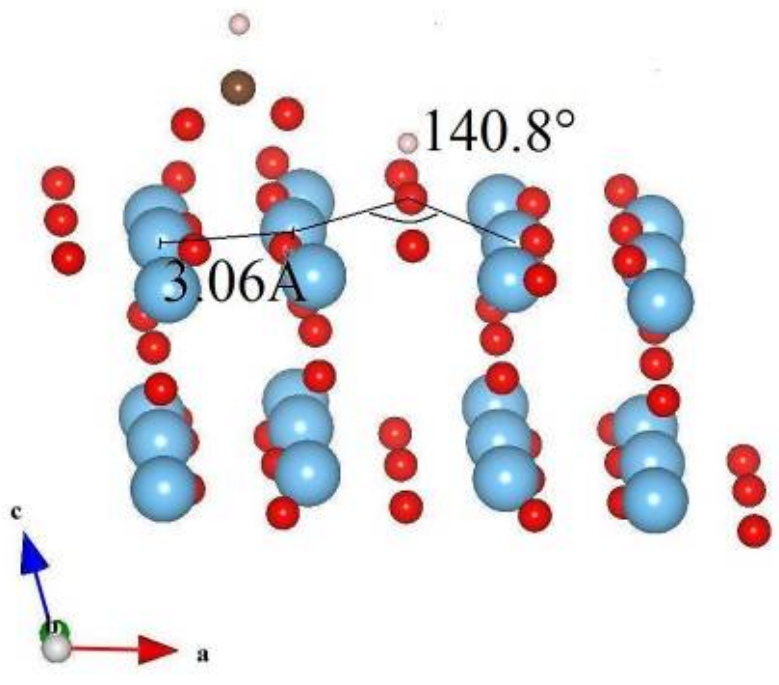

**Figure 7**

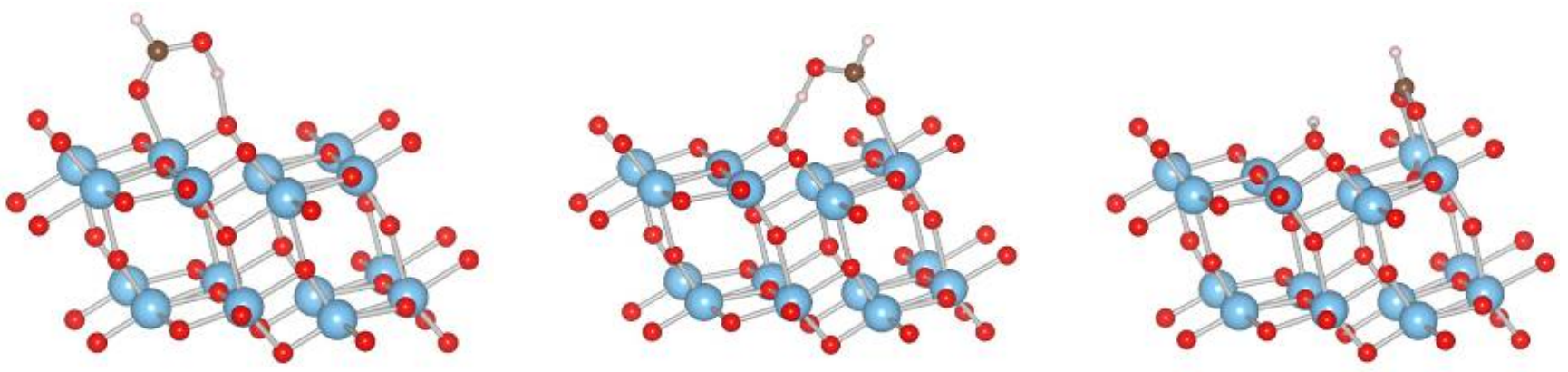

MHa_1 MHa_2 BB_H

**Figura 8**

| | Basis Set | a (Å) | b (Å) | c (Å) | β (°) |
|---|---|---|---|---|---|
| Anatase | PW | 3.818 | 3.818 | 9.617 | - |
| | LCAO | 3.790 | 3.790 | 9.629 | - |
| | Exp [1] | 3.784 | 3.784 | 9.515 | |
| Rutile | PW | 4.594 | 4.594 | 2.959 | - |
| | LCAO | 4.595 | 4.595 | 2.965 | - |
| | Exp [1] | 4.593 | 4.593 | 2.959 | - |
| $TiO_2$(B) | PW | 12.302 | 3.763 | 6.632 | 106.97 |
| | LCAO | 12.206 | 3.732 | 6.557 | 107.12 |
| | Exp [9,27] | 12.179 | 3.741 | 6.525 | 107.10 |

**Table 1**

| | LCAO-GGA | PW-GGA | PW-mBJ | Exp. (eV) |
|---|---|---|---|---|
| Rutile | 2.2 | 2.1 | 2.7 | 3.0 [1] |
| Anatase | 2.4 | 2.3 | 3.0 | 3.2 [1] |
| $TiO_2$(B) | 3.1 | 3.1 | 3.4 | 3.22 [27,28] |

**Table 2**

| Slabs | # of Ti | $TiO_2$ layers | SE ($J/m^2$) LCAO | SE ($J/m^2$) PW |
|---|---|---|---|---|
| A(101) | 8 | 2 | 0.31 | 0.40 |
| A(100) | 12 | 2 | 0.57 | 0.64 |
| A(001) | 5 | 4 | 0.89 | 0.90 |
| R(110) | 6 | 3 | 0.70 | 0.77 |
| R(101) | 6 | 3 | 1.09 | 1.14 |
| R(100) | 27 | 3 | 0.76 | 0.77 |
| TB(001) | 24 | 2 | 0.35 | 0.33 |
| TB(100) | 8 | 4 | 0.67 | 0.69 |

**Table 3**

| Slab | SE ($J/m^2$) SIESTA | Unsaturation (bb/$Å^2$) |
|---|---|---|
| A101 | 0.31 | 0.103 |
| A100 | 0.57 | 0.111 |
| A001 | 0.89 | 0.279 |

**Table 4**

| Slab | SE ($J/m^2$) SIESTA | Unsaturation (bb/$Å^2$) |
|---|---|---|
| R110 | 0.70 | 0.106 |
| R100 | 0.76 | 0.149 |
| R101 | 1.09 | 0.161 |

**Table 5**

| Slab | SE ($J/m^2$) SIESTA | Unsaturation (bb/$Å^2$) |
|---|---|---|
| TB001 | 0.35 | 0.0879 |
| TB100 | 0.67 | 0.0819 |

**Table 6**

| **Config.** | **Fig.** | **$E_{ads}$ (eV)** | **C-O1 (Å)** | **C-O2 (Å)** | **$\theta$ (°)** | **O1-Ti(5c) (Å)** | **O2-Ti(5c) (Å)** | **H-O(2c) (Å)** |
|---|---|---|---|---|---|---|---|---|
| MHa | 3 | **0.60** | 1.31 | 1.25 | 125.6 | --- | 2.23 | 1.51 |
| MHb | 3 | **0.15** | 1.38 | 1.21 | 123.4 | 2.40 | --- | 2.52* |
| MHa_3c | 4 | **0.55** | 1.31 | 1.25 | 125.7 | --- | 2.22 | 1.47 |
| M_H_2 | 3 | **0.02** | 1.35 | 1.22 | 124.3 | 1.94 | --- | 0.99 |
| BB_H_1 | 3 | **0.76** | 1.28 | 1.27 | 128.0 | 2.12 | 2.09 | 0.99 |

* This distance is much larger than the conventional hydrogen bond length (1.97Å)

**Table 7**

| Config. | Fig. | $E_{ads}$ (eV) | C-O1 (Å) | C-O2 (Å) | $\theta$ (°) | O1-Ti(5c) (Å) | O2-Ti(5c) (Å) | H-O(2c) (Å) |
|---|---|---|---|---|---|---|---|---|
| MHa_1 | 5 | **0.81** | 1.30 | 1.26 | 126.2 | --- | 2.13 | 1.49 |
| MHa_2 | 5 | **0.19** | 1.35 | 1.23 | 126.2 | --- | 3.61*[2] | 1.71 |
| MHb_1 | 5 | **0.38** | 1.40 | 1.21 | 122.0 | 2.28 | --- | 2.38*[3] |
| BB_H_1 | 5 | **1.06** | 1.28 | 1.27 | 128.6 | 2.06 | 2.05 | 0.98 |
| BB_H_2 | 5 | **1.35** | 1.29 | 1.27 | 128.8 | 2.02 | 2.06 | 0.98 |
| M_H_1 | 5 | **0.40** | 1.37 | 1.22 | 122.7 | 1.89 | --- | 1.00 |
| M_H_2 | 5 | **0.69** | 1.38 | 1.22 | 122.3 | 1.84 | --- | 0.98 |

*[2] This does not represent a typical Ti-O bond length.

*[3] This distance is much larger than the conventional hydrogen bond length (1.97Å)

**Table 8**

| Config. | C-O1 (Å) | C-O2 (Å) | $\theta$ (°) |
|---|---|---|---|
| TB001_L | 1.37 | 1.22 | 125.4 |
| TB100_L | 1.37 | 1.22 | 125.2 |

**Table 9**

| Config. | Fig. | $E_{ads}$ LCAO | $E_{ads}$ LCAO_FB | $E_{ads}$ PW |
|---|---|---|---|---|
| MHa_1 | 8 | -0.68 | -0.82 | -0.75 |
| MHa_2 | 8 | -0.63 | -0.88 | -0.76 |
| BB_H | 8 | -0.28 | -0.65 | -0.56 |

**Table 10**